\documentclass[twocolumn]{article}
\usepackage{times,graphicx,color}
\usepackage{epsfig}

\begin{document} 
\bibliographystyle{pnas1}
\title{\textbf{\LARGE Oscillation patterns in negative feedback loops}}
\author{Simone Pigolotti$^1$, Sandeep Krishna$^2$, Mogens H. Jensen$^2$\\
$^1$IMEDEA, C/ Miquel Marqu\`es, 21, 07190 Esporles, Mallorca, Spain.\\
$^2$Niels Bohr Institute, Blegdamsvej 17, DK-2100 Copenhagen, Denmark.}
\date{}

\maketitle 

\abstract{Organisms are equipped with regulatory systems that
  display a variety of dynamical behaviours ranging from simple stable steady
  states, to switching and multistability, to oscillations.  Earlier work
  has shown that oscillations in protein concentrations or gene expression
  levels are related to the presence of at least one negative feedback loop in
  the regulatory network.  Here we study the dynamics of a very general class
  of negative feedback loops. Our main result is that in these systems the
  sequence of maxima and minima of the concentrations is uniquely determined by
  the topology of the loop and the activating/repressing nature of the
  interaction between pairs of variables.  This allows us to devise an algorithm
  to reconstruct the
  topology of oscillating negative feedback loops from their time
  series; this method applies even when some variables are missing from the
  data set, or if the time series shows transients, like damped oscillations.
  We illustrate the relevance and the limits of validity of our method with
  three examples: p53-Mdm2 oscillations, circadian gene expression in
  cyanobacteria, and cyclic binding of cofactors at the estrogen-sensitive pS2
  promoter. }

\section*{Introduction}

Physiological processes in living cells exhibit a wide range of dynamical
behaviour.  Response systems which regulate the levels of potentially
poisonous substances like iron 
\cite{MA}, or other stresses like DNA damage 
\cite{Friedmanetal} are typically homeostatic, meaning that
the concentrations and expression levels of the involved proteins and genes
eventually return to a stable, constant level after an external perturbation.  Other
systems are designed to be multistable: $\lambda$ phage upon entering a host
bacterium has a core regulatory network that puts it into either a virulent,
lytic state or a dormant, lysogenic state \cite{lphage}.  The dynamics of
such systems is similar to those having a single stable steady state except
that noise and external perturbations can cause switching between states. For
example, damage to the bacterial genome can cause the state of the $\lambda$
phage to switch from lysogenic to lytic \cite{lphage}.  A third class of
subsystems consists of those that exhibit oscillations. The most obvious
examples are cell division and circadian (24 hour) rhythms.  Cellular
processes are often coupled to the circadian clock, e.g. respiration and
carbohydrate synthesis in cyanobacteria \cite{GIJK}, which makes them periodic
also.  Recently several systems of interacting proteins have been found
showing faster, ``ultradian" oscillations with time periods of the order of
hours, which influence the immune system (NF-$\kappa$B \cite{HLSB,Nelsonetal}),
apoptosis (p53 \cite{alonp53}), and development (Hes \cite{Hirataetal}).

Theoretical studies of these oscillatory systems
\cite{Goldbeter_circadian,nfkb,mogenstiana,mogensHes} usually describe the
dynamics with differential equations modeling the following kinds of
interactions: regulated gene transcription and translation, active or passive
degradation of proteins and mRNA, protein modifications (e.g.,
phosphorylation, methylation), complex formation and transport processes.
Among these, transcription, translation and degradation are the ones which
occur on the slowest timescales -- minutes to hours. Complex formation and
protein modifications are typically much faster, on the order of seconds.
Transport processes, even if they are actively catalyzed by molecular motors
or pumps, are often slower than such chemical reactions. Thus, if one is
interested in the dynamical behaviour only over long timescales, the models
can be simplified by averaging over the faster processes
\cite{celldivisionPRE}.  This coarse grained description should be adopted
with care, since some biochemical processes have recently been found to
display oscillations on their own, even on very long (circadian) timescales
\cite{nakajimacircadian,doubttranscription}.  

Mathematically, this simplification is very useful because the slower
processes mentioned above, regulated transcription and translation,
degradation and transport, are usually {\sl monotone}: when the concentration
of one chemical is changed, the qualitative effect on the other species (an
increase or decrease of their concentration) is independent of the
concentrations of the chemicals.  In other words, in general, proteins
that activate a particular process will not change to repress that process
at a different concentration.
Indeed, an important conjecture by Thomas \cite{thomas},
rigorously proven in refs. \cite{snoussi, gouze}, states that in a monotone
system at least one positive feedback loop is needed in order to have
multistability (i.e., existence of multiple steady states), and at least one
negative feedback loop is needed in order to have periodic behavior.

Feedback loops may thus be seen as the building blocks of the nontrivial dynamical
behaviors of these systems; a network without loops can only reach a unique
fixed point, regardless of the initial conditions. A deeper 
understanding of the dynamics of these simple units would help us gain
insight into the dynamics of more complex and structured biological networks.
These more complicated models of oscillating biological systems can be built
up not only by considering multiple, intertwined loops, but also by introducing
time-delays \cite{mogenstiana,mogensHes}, noise \cite{forger} or spatial
effects \cite{Goldbeter_cAMPspirals} in the dynamics: all these effects
greatly enhance the range of dynamical behaviors of these systems. However, a
drawback is that discriminating between these possibilities
(and selecting the most reasonable model accordingly) often demands 
high precision data and long time series 
that may be difficult to obtain experimentally.

\begin{figure}[tb]
(a)\\
\vspace*{-0.5cm}
\begin{center}
\includegraphics[width=8.5cm]{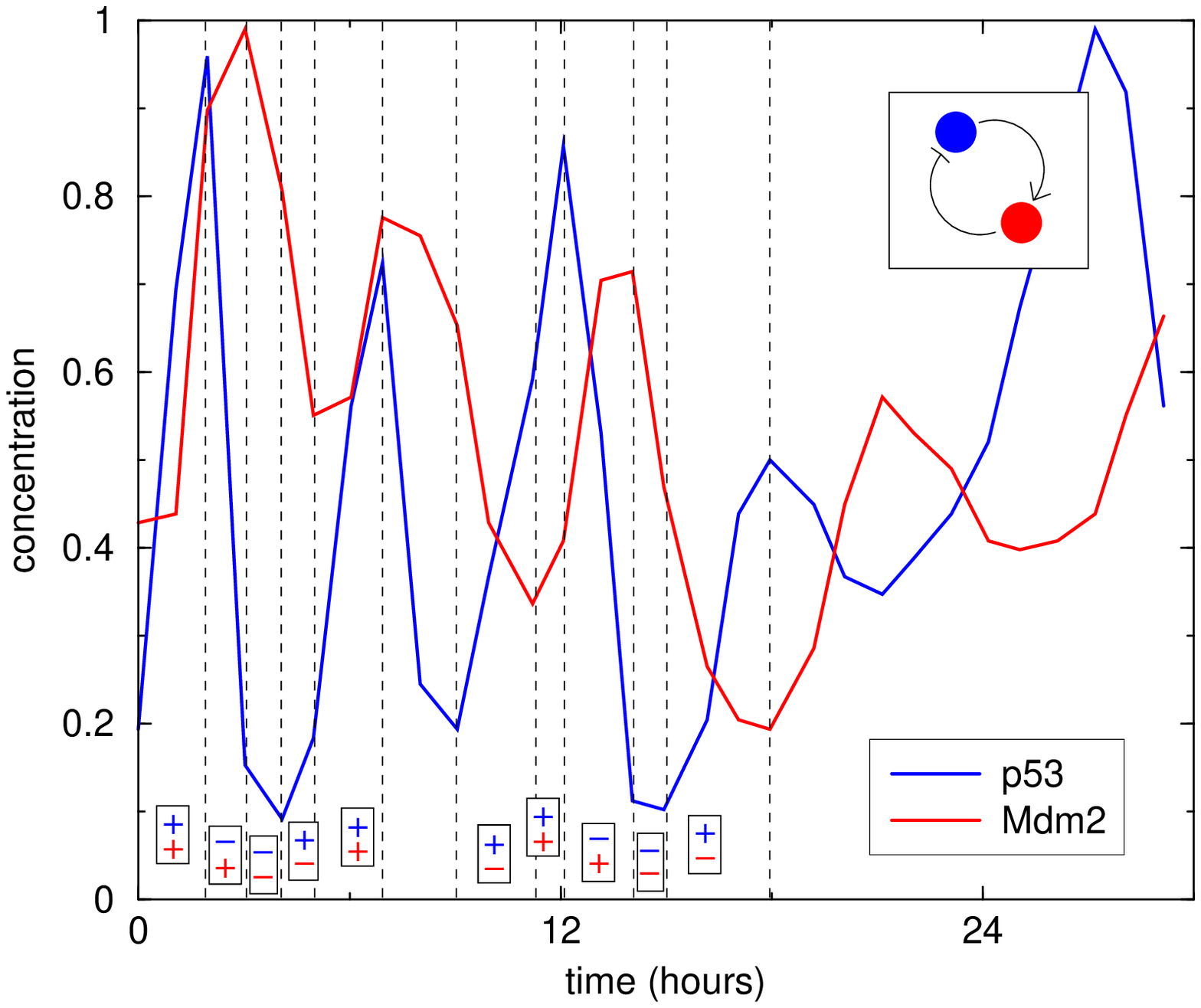}
\end{center}
(b)\\
\vspace*{-0.5cm}
\begin{center}
\includegraphics[width=7cm]{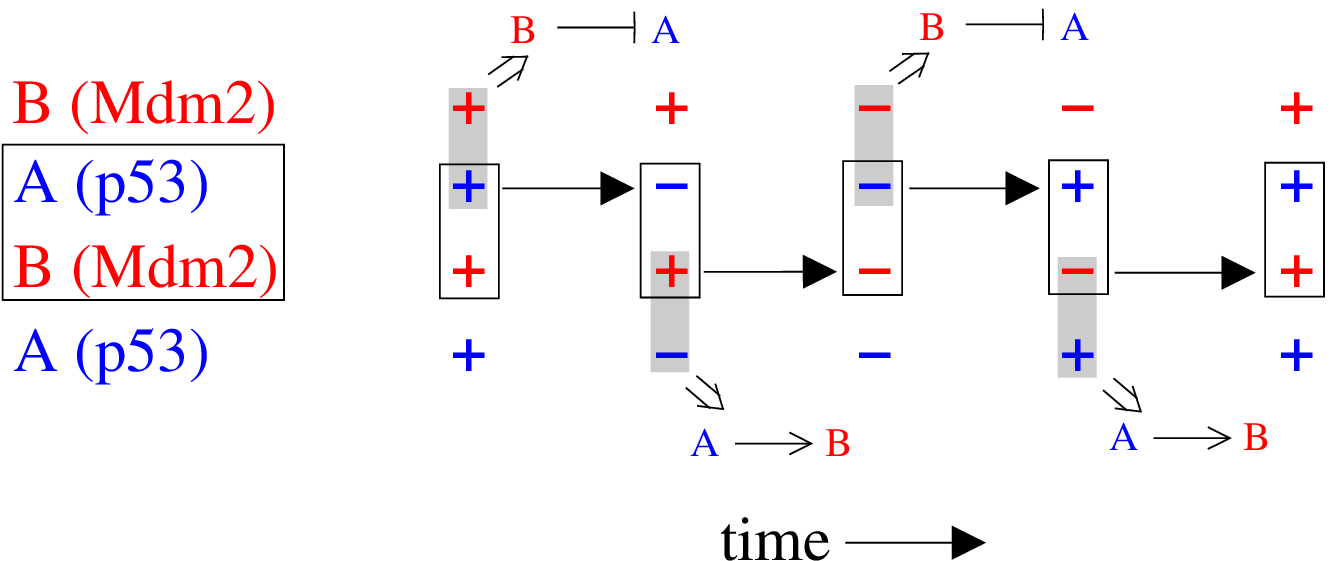}
\end{center}
\caption{(a) p53-Mdm2 oscillations as recorded in a fluorescence microscopy experiment 
  \cite{alonp53} and the reconstructed symbolic dynamics.
Inset shows the negative feedback loop extracted using the algorithm in the text;
the process is shown in (b). 
Here, and in subsequent figures, ordinary arrows represent activation, while barred arrows represent
repression.\label{figurep53}}
\end{figure}

As an example, consider the p53-Mdmd2 oscillations observed in single-cell
fluorescence experiments, shown in Fig. \ref{figurep53}a.  One could ask, what
causes the fluctuations in the amplitude of the observed oscillations? Are
they caused by an underlying chaotic dynamics, by noise due to interaction
with other proteins or by time-delay effects?  Even if there are rigorous
time-series techniques to answer to this question \cite{kantz}, they are
unlikely to give a conclusive answer because of the low statistics.

Nevertheless, the oscillations in Fig. \ref{figurep53}a contain precious
information about the real system: notice that the time order of the maxima
and minima of the concentrations is always the same (where it is possible to 
discriminate the sequence).  In this paper we show
that these sequences have a precise pattern for a large class of
deterministic, non-time-delayed models of negative feedback loops, meaning
that they can be used to deduce the precise order of activators and repressors
in the feedback loop generating the time series.  We have devised a simple
algorithm for doing this which
is described in the next section. The mathematical basis for the algorithm
is laid out in the subsequent two sections.
In the final section, we apply our
algorithm to reconstruct the feedback loop from oscillating time series of
two more biological systems; the examples also clarify the the scope
and limitations of the algorithm.

\section*{Extracting the feedback loop}

As mentioned above, the time series in Fig. \ref{figurep53}a shows a fixed order
of maxima and minima of the concentrations.
We can investigate this pattern by
dividing the dataset into intervals whose ends are determined by the occurrence
of an extremal value (maximum or minimum) of a variable. In each of these
intervals, marked by vertical dotted lines in Fig. \ref{figurep53}a, each
variable shows an unchanging trend, either increasing or decreasing with time.
We can therefore uniquely associate to each interval a ``symbol" of the form
$(+,-,+,\ldots)$, containing one sign for each variable, with a `$+$' meaning
that that variable is increasing and a `$-$' meaning it is decreasing.  In Fig. 1,
each box corresponds to one such symbol (for convenience the signs are arranged
vertically in all figures, but horizontally in the text). Thus,
the continuous time series is converted into a discrete sequence of symbols,
which we term the ``symbolic dynamics" \cite{Strogatzbook}.  The algorithm listed below can then
determine whether the sequence is consistent with the dynamics of a single
negative feedback loop, and if so, the precise order of activators and
repressors in that loop.\\

{\bf The algorithm}
\begin{enumerate}
\item List the order in which the maxima and minima of the variables occur. E.g., in
Fig. \ref{figurep53}a, the order is p53 max, Mdm2 max, p53 min, Mdmd2 min, p53 max, Mdm2 max ...
\item If the variables occur in this list in an unchanging cyclic order, then this fixes the
order of species in the loop, i.e., a variable activates or represses the one immediately
following it in the list. Otherwise, 
a single negative feedback loop is inconsistent with the time series. 
\item Construct the symbolic dynamics for the time series, with +/- symbols listed in the
order obtained in step 2.
\item If the symbolic dynamics is not periodic,
a single negative feedback loop is inconsistent with the time series. Otherwise, start with
any variable and the one pointing to it (say, variables B and A) and note the steps
where B changes sign. If at the previous step (before B changed), A had the same sign as B 
then A represses B, else it activates B (see Fig. \ref{figurep53}b).
\item This procedure is repeated for each variable to obtain the effect of the preceding variable.
E.g., in Fig. \ref{figurep53} we conclude that p53 activates Mdm2, and Mdm2 represses p53.
If the various sign changes of any variable give inconsistent conclusions,
then a single negative feedback loop is inconsistent with
the time series. 
\item If the number of repressors in the loop is even, then a single negative
feedback loop is inconsistent with the time series.
\end{enumerate}

For p53-Mdm2 oscillations, there are only two possibilities for a single
negative feedback loop involving only these two. Either p53 activates Mdm2
which represses p53, or p53 represses Mdm2 which activates p53.  The above
algorithm applied to Fig. \ref{figurep53}a picks out the former, for which
experimental evidence already existed \cite{Wuetal} (discussed further in the ``Examples" section).  This is a particularly
simple case, as it involves only two variables. Our algorithm aids model
selection much more when there are more than two variables involved, as shown 
in the two other examples discussed in the final section of the paper. First, we
establish the mathematical basis for this algorithm.

\begin{figure}[t]
\begin{center}
\includegraphics[width=9cm]{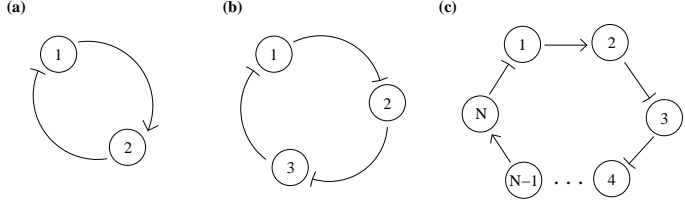}
\end{center}
\caption{Examples of negative feedback loops: a) The simplest case consisting of one activator and
one repressor. 
b) a 3-repressor loop, c) a general loop with $N$ variables and an odd number of repressors.
\label{schematicloops}} 
\end{figure}

\section*{A general class of negative feedback loops}
Consider a feedback loop composed of an arbitrary number, $N$, of nodes, where
each node can be either an activator or a repressor (see Fig. \ref{schematicloops}c).  
Nodes could be genes, proteins, metabolites or any other chemical species
which could, directly or indirectly, activate or repress other nodes in the system.
The equations we study are of the form:
\begin{equation}
\label{eq:model}
\frac{dx_i}{dt}=g_i^{(A,R)}(x_i, x_{i-1})\qquad i=1\ldots N.
\end{equation}

$x_i$ is a dynamical variable associated to node $i$, e.g., the concentration
of the chemical species $i$, or the expression level of gene $i$. Henceforth,
we will call $i$ a chemical species and $x_i$ its concentration.  The vector
field, whose components are $g_i(x,y)$, contains all the basal production,
degradation, and possibly self-catalytic terms, and specifies the interaction
between variables. We denote explicitly with the superscript that each species
$i$ is either activated (A) or repressed (R) by the species $i-1$ immediately
preceding it in the loop.  We allow for heterogeneity, meaning that the
different species can be characterized by different production and degradation
rates, and different interaction strengths, i.e., different functions $g_i$
for each $i$. The functions can also be nonlinear. For instance, one way of implementing the 
3-repressor loop
in Fig. \ref{schematicloops}b would be:
\begin{equation}\label{simpleexample}
\frac{dx_i}{dt}=c-\gamma x_i+\alpha\frac{1}{1+(x_{i-1}/K)^h},
\label{3cycle_eq}
\end{equation}
for $i=1,2,3$ (with $i=0$ the same as $i=3$). These equations model
the basal production of each protein ($c$), the uniform dilution of
each proteins by cell growth ($\gamma$) and the production of each protein
($\alpha$) that is repressed 
by the one behind it in the circuit.
The repression we have chosen to be of a standard
Michaelis-Menten form, with half-maximum at $K$. The Hill coefficient,
$h$, models the cooperative prevention of transcription
by $h$ molecules of $i-1$ binding at the promoter of $i$.
This is a simplified version of the repressilator \cite{repressilator}.

This example simply provides a single illustration of possible
$g_i$ functions. In fact, we do not constrain the $g_i$ to be
like eq. \ref{3cycle_eq}.
The only restrictions we put on the
functions $g_i$ are the following two conditions:

\begin{enumerate}
\item All trajectories should be bounded and persistent, meaning that
  all the concentrations should stay positive and finite in the time
  evolution.
\item Monotonicity: all the $g_i(x,y)$s have to be monotonically decreasing
functions of the first argument. Moreover, the $g^R(x,y)$ are decreasing
functions of the second argument, while the $g^A(x,y)$ are increasing
functions of the second argument.
\end{enumerate}
Condition 1 is imposed to ensure that concentrations of the species are well
defined and cannot grow infinitely, a biologically plausible constraint: typically
for regulatory networks, $g_i$ is bounded above (i.e., there is a maximum
possible rate of production) and is dominated by the degradation terms for
sufficiently large concentrations, thus ensuring condition 1.  Condition 2
implies that activators of a specific process are activators at all concentrations (and
similarly for repressors).  In other words, we exclude regulation like CI in
lambda phage which can activate the $P_{\mathrm{RM}}$ promoter at low concentrations, but repress it at high
concentrations 
\cite{CIregulation2}.  Another example is the galactose
regulator GalR, which at high concentrations of galactose is an activator of
the promoter {\sl galP2} but in the absence of galactose 
forms a DNA loop, in which state {\sl galP2} is completely repressed
\cite{semseyvirnik}.  Such examples are, however, relatively rare and we
exclude them from the class of networks we study.

The monotonicity condition can be used to prove that if the number of
repressors in the loop is even there can be multiple fixed points. This is
necessary for bistable or multistable systems, as previously analyzed in
\cite{pnasangeli}. In the following we focus on the case of an odd number of
repressors, i.e., negative feedback. Then, as shown in
supplementary material, there is one and only one fixed point.  Further, a
linear stability analysis shows that the transition to instability is
necessarily a Hopf bifurcation, which implies that near the transition point
there exists a periodic orbit (see supplementary material).
However, this stability analysis allows one to study the dynamics only
locally, both in the phase space, i.e., close to the fixed point, and in
parameter space, i.e., close to the bifurcation value. In the next section, we
show that in general, whether the fixed point is stable or unstable, there are
restrictions on the trajectory of the system.

\section*{Symbolic dynamics}

Our argument is the direct consequence of how the nullclines, i.e., the $N$
manifolds defined by the equations $g_i(x_i,x_{i-1})=0$, partition the phase
space (the positive orthant $x_i>0 \ \ \forall i$.) Each nullcline separates
two fully connected regions, one in which the $i$th component of the field,
$g_i$, is positive and one in which it is negative. Furthermore, all these
manifolds have exactly one point in common, the unique fixed point ${\mathbf
x}^*$. The phase space is thus the union of $2^N$ sectors, each characterized
by the signs of the components of the field.  These sectors correspond
precisely to the symbols $(+,-,-,+,\ldots,-)$ defined previously.

Notice that there cannot be an attractor entirely contained in the interior of
a sector because the field does not change sign, the trajectory is bounded and
there is only one fixed point.  Thus, either the trajectory of the system
spirals in towards a stable fixed point, or, if the fixed point is unstable,
it will keep crossing from one sector to another indefinitely. In the first
case, the symbolic dynamics is a finite sequence of symbols, which ends when
the trajectory stops crossing sector boundaries. In the latter case, it will
be an infinite sequence of symbols. In either case, adjacent symbols in the
sequence will differ by only one sign change.

\begin{figure}[tb]
\begin{center}
\includegraphics[width=8.5cm]{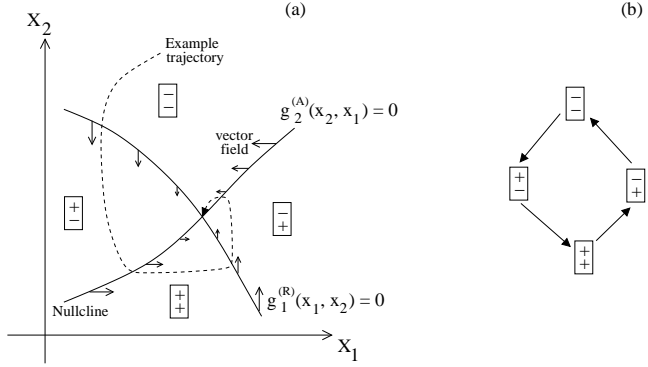}
\end{center}
\caption{(a) Schematic diagram of the phase-space of the 2-variable
negative feedback loop of Fig. \ref{schematicloops}a. Solid lines show
the two nullclines, which intersect and divide the space into four
sectors, labelled by the signs of $g_1$ (top sign) and $g_2$ (bottom) in that
sector. Arrows on the nullclines show the local direction of the
vector field, which determines the direction in which the nullcline
can be crossed by a trajectory.  The dotted line is an example
trajectory which follows these rules, while spiralling in towards the
stable fixed point.  (b) The allowed transitions for moving from one
sector to another. The symbolic dynamics of any trajectory has to be
consistent with these rules. \label{2dsymb}}
\end{figure}

The key point of our argument is that any of the boundaries can be
crossed in just one specific direction, due to the monotonicity of the
functions $g_i(x_i, x_{i-1})$. This means that not all possible sign
changes are allowed. In Fig. \ref{2dsymb}, we illustrate this using the
example of the 2-species negative feedback loop of
Fig. \ref{schematicloops}a, one repressor and one
activator. Fig. \ref{2dsymb}b shows the only 4 transitions possible
in this system. Thus, starting from any initial condition, the
symbolic dynamics will follow the order shown in Fig. \ref{2dsymb}b
until the trajectory converges to the stable fixed point and there are
no further sign changes. This example gives a good pictorial idea of
the fact that the nullclines behave like one-way doors.

In the general $N$-species case, the same phenomenon occurs, with the
symbolic dynamics obeying the following rules (see supplementary material for more information).
\begin{itemize}
\item If the variable $(i-1)$ represses $i$, the nullcline $i$
  can be crossed if $g_i$ and $g_{i-1}$ have the same sign.
\item If the variable $(i-1)$ activates $i$, the nullcline $i$
  can be crossed if $g_i$ and $g_{i-1}$ have opposite signs.
\end{itemize}
To emphasise the point further, the direction in which the nullcline $g_i=0$
can be crossed at a given point depends on the position of that point relative
to only one other nullcline, $g_{i-1}=0$. It does {\em not} depend on any
other nullcline.  In simple words, the rules encode the fact that an
increasing activator can make the affected concentration increase (but not
decrease), while an increasing repressor can make the affected concentration
decrease (but not increase).  Note that the allowed transitions apply to any
trajectory, even transients.  Thus, if one is analysing oscillatory time
series it doesn't matter whether the oscillations are sustained, or the
measurement is of transients or damped oscillations.

To determine the general scenario which is compatible with these
rules, consider that when a nullcline is crossed, the symbolic
dynamics makes a transition between two states differing by just a
single sign. We say that there is a {\em mismatch} between two
adjacent signs if the nullcline depending on these two variables can
be crossed according to the rules defined above. The effect of
crossing the nullcline $i$ is to remove the mismatch between $i$ and
$i-1$. If there was also a mismatch between $i$ and $i+1$, it too is
removed, otherwise it is created. For a negative feedback loop, we can
show that the number of mismatches has to be odd, and cannot increase.

Now consider what happens if the fixed point is unstable.
If there is just one mismatch, it can only keep traveling around the
loop, in the direction of the loop arrows.  This means that
the symbolic trajectory is periodic of period $N$. In the general case
we can visualise the symbolic dynamics as several mismatches traveling
around the loop in the same direction. Whenever two mismatches ``hit"
each other, they annihilate. Eventually the number of mismatches will
reach some limit, where each mismatch stays safely distant from the
others. The length of the loop limits how many mismatches can, in
principle, coexist; for example, only one mismatch can survive if
$N<4$. In practice, even in long loops, it is likely that only one
mismatch survives and we will restrict to this case in the following.

An interesting consequence of this periodicity is that any of the
$N$ nullclines defines a Poincar\'e map for the dynamical system.
Periodic oscillations in the symbolic dynamics translate into a stable
periodic orbit if the dynamics on the Poincar\'e map reaches a fixed
point, or into chaotic oscillations if the dynamics on the Poincar\'e
map is chaotic.

Whether the orbits are chaotic or not we have proven, for this general class
of systems, that when the fixed point is unstable, the dynamics is oscillatory
with well defined properties. For example, each of the concentrations has
exactly one maximum and one minimum during a time period of the symbolic
dynamics, and the fact that the mismatch travels in the direction of the
feedback loop implies that the sequence of maxima and minima has to follow the
order of the species in the loop. From the particular order observed it is
also possible to argue which species acts as an activator and which as a
repressor. Furthermore, the observation of a time series which is incompatible
with the symbolic dynamics rules allows one to exclude a dynamics of the form
of eq. (\ref{eq:model}), generally suggesting a topology more complicated than a
simple feedback loop or more subtle effects like time delays and non-monotonic
regulation.  The algorithm described previously is a straightforward
consequence of these rules.

Notice that our method works even if one does not measure the time series of all
the species belonging to the loop. The algorithm gives a coherent conclusion
about the overall sign between the variables: for example, a variable A will
appear as an activator of a variable B if there is an even number of
``unobserved'' repressing links between them (see supplementary material).
The following examples will further clarify these points.

\section*{Examples}

We now apply the above ideas to extract information about the loop
structure from experimentally observed time series of three systems: 
p53-Mdm2 oscillations in mammalian cells, circadian expression 
of {\em kai} genes in {\em Synechocystis} cyanobacteria, and cyclic binding
of protein cofactors with DNA at the estrogen-sensitive pS2 promoter in human breast cancer cells.

Our first example is the well-known p53-Mdm2 negative feedback loop, already
discussed in the introduction.  The tumor suppressor protein, p53, is a
transcription factor that affects the expression of a large number of genes
including several involved in cell cycle arrest and apoptosis, while Mdm2 is
an important regulator of p53 activity \cite{p53levinereview}.  For the
oscillating concentrations of p53 and Mdm2 in Fig. \ref{figurep53}, in a
couple of cases there is some ambiguity about the order of maxima and
minima. However, the pattern is clarified by comparing two separate time
intervals in which the symbolic sequence is unambiguous.  Notice that both
regions exhibit the same periodic symbolic sequence. According to the rules we
stated in the previous section, this sequence is allowed if p53 activates Mdm2
and Mdm2 represses p53. This is completely consistent with independent
experimental knowledge of the system. p53 activates transcription of the {\em
mdm2} gene 
\cite{Wuetal}.  Mdm2, once produced, binds to p53
preventing it from acting as a transcription factor, and subsequently
ubiquitinates it, which enhances its proteolytic breakdown
\cite{Wuetal}.  Thus, Mdm2 is a
repressor of p53. Therefore, we conclude: (i) the observed oscillations are
consistent with a dynamics of the form of eq. (\ref{eq:model}), (ii) however, for
this there must be at least one other unobserved species taking part in the
loop, since the fixed point is always stable for $N=2$. Indeed, several three
variable models of p53-Mdmd2 oscillations have been examined, which assume the
third variable to be either an Mdm2 precurser (e.g., Mdm2 mRNA) or a third
protein which interacts with p53 or Mdm2 \cite{alonp53}.  Of course, it is
hard to say at this level if a simple, deterministic negative feedback loop is
a good model for this specific system. For example, the same sequences of
symbols could be observed in a time-delay model \cite{mogenstiana}. In
addition, experiments suggest the presence of at least 10 feedback loops
involving p53 \cite{p53levineoscillations}.  It is also nontrivial to assess
whether the aperiodicity of the trajectory is due to internal mechanisms
(chaos, time delays), or to interaction with other proteins. Still, we can
conclude that a model like eq. (\ref{eq:model}) is a good candidate for a
zero-order model, being simple and reproducing correctly the qualitative
behavior of the components with the correct interaction signs.

\begin{figure}[h!]
\begin{center}
(a)\hfill~\\
\includegraphics[width=8.5cm]{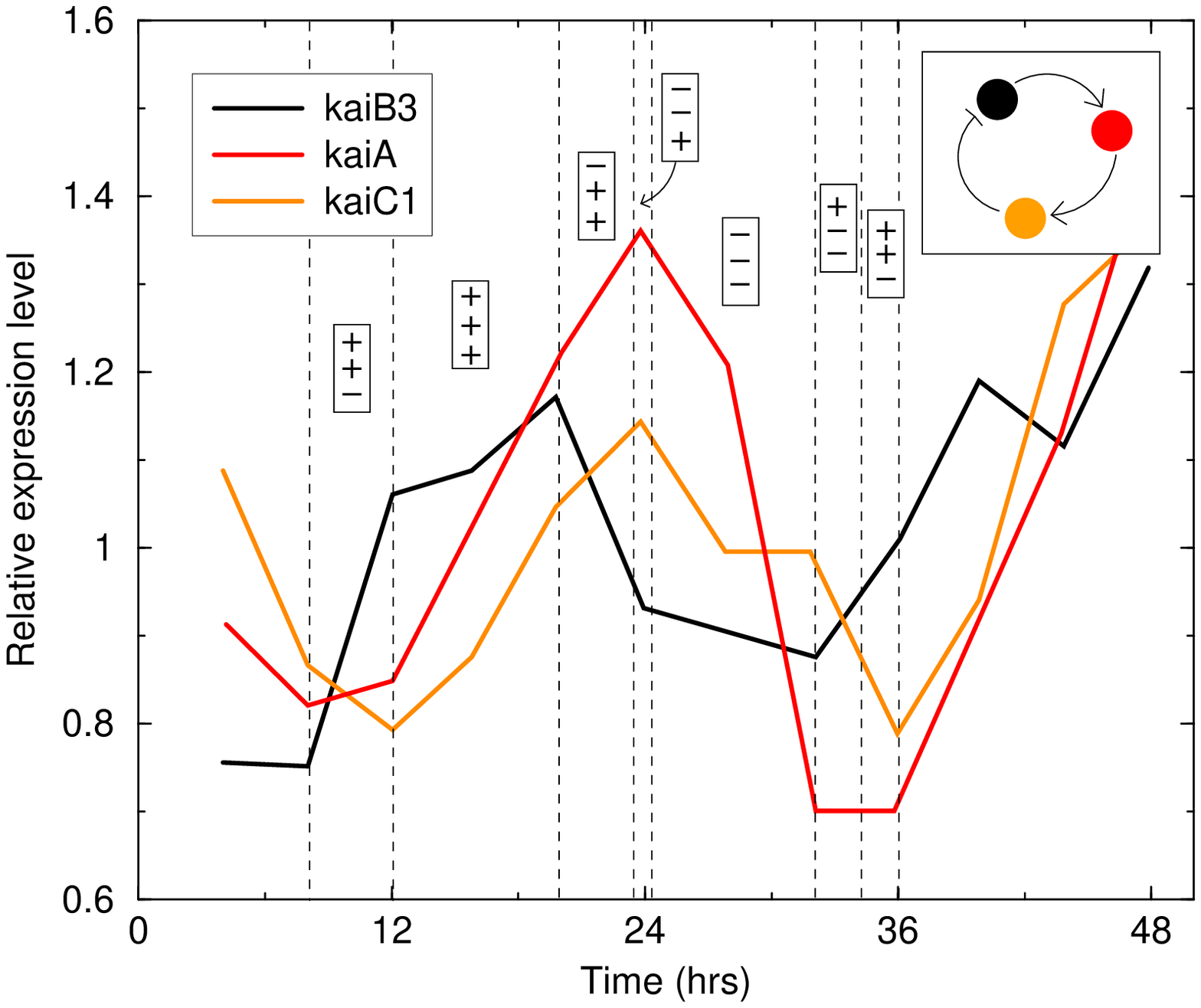}\\
(b)\hfill~\\
\includegraphics[width=8.5cm]{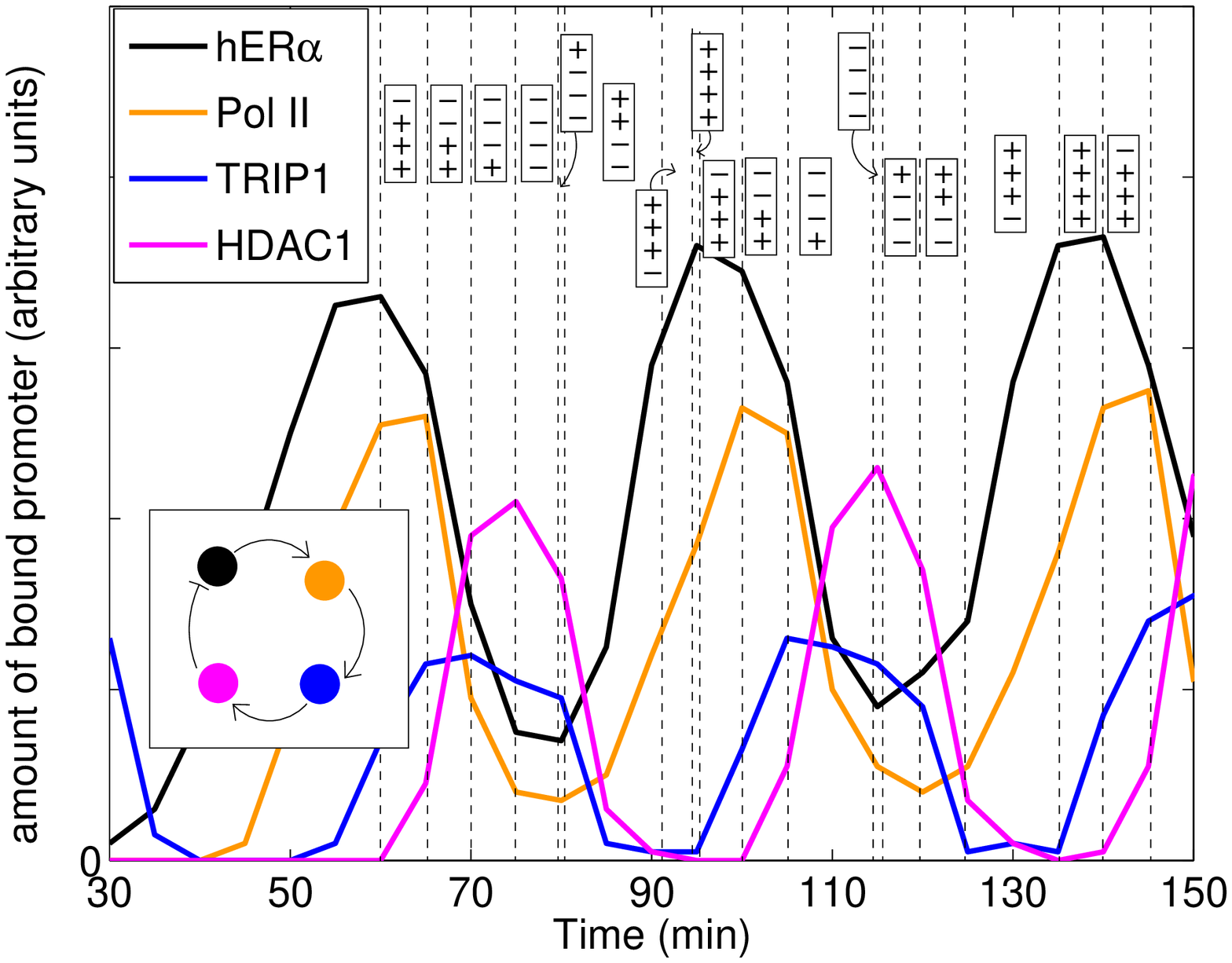}\\
\end{center}
\caption{(a) Circadian rhythms of three {\em kai} genes 
in a {\em Synechocystis} cyanobacterial strain (data from ref. \cite{kuchocircadian}).
(b) Periodic binding of four proteins to the pS2 promoter following
addition of estradiol (data from ref. \cite{lemaireprl}, based on ref. \cite{meticell}). In each case the corresponding symbolic
dynamics is also shown, with symbols in the same order as the legend (where maxima/minima of two variables occur very
close we have exaggerated the separation between the dotted lines for visual clarity). 
The insets shows the topology deduced from the symbolic dynamics.
\label{figurecyanops2}}
\end{figure}

The next example involves circadian oscillations of gene expression in
cyanobacteria. Cyanobacteria are the only bacterial species with a circadian
clock and several of their cellular functions appear to be under circadian
control \cite{GIJK}. In {\em Synechococcus elongatus} a cluster of
three genes, {\em kaiA,B,C}, were found to be essential: deletion of any of
these genes eliminated the oscillations \cite{Golden_JohnsonKondo}.  
Fig. \ref{figurecyanops2}a shows circadian rhythms in the expression levels
of genes coding for homologues of the KaiA,B,C proteins in one {\em Synechocystis} strain.
The symbolic dynamics is
consistent with a three-variable feedback loop (Fig. \ref{figurecyanops2}a,
inset), where {\em kaiA} activates {\em kaiC1}, which represses {\em
kaiB3}, which, in turn, activates {\em kaiA}.  The first two of these
predicted interactions exist in {\em Synechococcus}
\cite{Golden_JohnsonKondo}, while the third is a new prediction for how {\em
kaiA} is brought into the loop. Note that our analysis only provides the sign
of this interaction. It does not reveal the molecular mechanism of the
interaction, nor whether the interaction is direct or through
intermediate steps.

Finally, we consider the cyclic binding of cofactors to the estrogen-sensitive
pS2 promoter. A coordinated sequence of binding and unbinding events modifies
the DNA packing and nucleosome structure to enable transcription to proceed
\cite{meticell}.  This is a case where no model exists and not all the
proteins involved have been identified.  Our method is particularly suited for
such a case, because it does not matter if the dynamics of only a subset of
the proteins involved is available.  Ref. \cite{meticell} measured the
temporal dynamics of binding of several proteins at the pS2 promoter using ChIP assays.
Fig. \ref{figurecyanops2}b shows oscillations in the binding of 4 proteins,
after the addition of estradiol.  ER, estradiol receptor, binds estradiol and
is required for initiating transcription.  Pol II is the RNA polymerase that
transcribes the gene. TRIP1 is a component of the APIS proteasome subunit, while HDAC1 
is involved in deacetylation of histones \cite{meticell}.  The
symbolic dynamics is consistent with the model shown in
Fig. \ref{figurecyanops2}b, inset.  Notice that in this case, each variable
measures the amount of bound protein at the pS2 promoter (albeit in arbitrary units,
which are different for each protein).
The predicted links indicate how a bound protein affects the probability of binding (or
of remaining bound) of another one in the sequence. For example, the link from
ER to Pol II indicates that ER, when bound at the promoter, increases the
recruitment probability of Pol II. Ref. \cite{lemaireprl} models the dynamics
of Fig. \ref{figurecyanops2}b using a {\it positive} feedback loop, requiring
over 200 intermediate steps, which has only activating links.  Our analysis
predicts the existence of a repressive link between HDAC1 and ER. One could
imagine this repression as a result of a change in the conformational
state of the DNA or histones, or the blocking of a binding site for ER at the
promoter. However, we emphasise again that our analysis does not give any
information about the molecular mechanism for this repression or whether there
are intermediate steps.  The analysis does suggest that, for this system, a
negative feedback loop is a plausible hypothesis as the cause of
oscillations.

\section*{Acknowledgements}
We thank Ian Dodd for critical reading of the manuscript and many useful suggestions. This work was supported
by the Danish National Research Foundation.

\section*{Supplementary Material for ``Oscillation patterns in negative feedback loops"}

\section*{Fixed point analysis}
In this section we study the fixed point properties of 
a feedback loop composed of an arbitrary number, $N$, of nodes
whose dynamics is given by Eq. (1) in the main text, which we repeat here:
\begin{equation}
\label{eq:model_s}
\frac{dx_i}{dt}=g_i^{(A,R)}(x_i, x_{i-1})\qquad i=1\ldots N.
\end{equation}

Our analysis proceeds by noting that, using the monotonicity
condition, we can write explicit functional relations between
neighboring variables in the steady state (when $dx_i/dt=0$):
                                                                                                                                                                                                                                             
\begin{equation}
g_i^{(A,R)}(x_i^*,x_{i-1}^*)=0\quad\Rightarrow\quad  x_i^*=f_i^{(A,R)}(x_{i-1}^*)
\end{equation}
Notice that the functions $f_i$ have the same monotonicity
properties as the $g_i$s with respect to the second
argument (for this it is necessary that $g_i(x,y)$ be a monotonically {\em decreasing} function of $x$). 
By iterative substitution, we obtain:
\begin{eqnarray}\label{fixed2_s}
x_i^*=f_i(x^*_{i-1})=f_i(f_{i-1}(x^*_{i-2}))=\ldots=\nonumber\\
=f_i\circ f_{i-1}\circ f_{i-2}\circ\ldots\circ f_{i+1}(x^*_i)\equiv F_i(x^*_i)
\end{eqnarray}
where $\circ$ denotes convolution of functions. Here, we introduced
the function $F_i(x)$, which quantifies how the species $i$ interacts
with itself by transmitting signals along the loop.  Notice also that
if Eq.(\ref{fixed2_s}) holds for one value of $i$, then it holds for any
$i$, since it is sufficient to apply $f_{i+1}()$ on both sides to
obtain the equation for $x_{i+1}^*$ and so on. For feedback loops,
much useful information can be obtained from the properties of
$F_i(x)$.  Firstly, by applying the chain rule, we obtain the slope of
$F_i(x)$ at $x$:
$F_i'(x)=\prod_j f_j'(x_j)|_{x_i=x}$.  The r.h.s is always greater
(less) than zero if the number of repressors present in the loop is
even (odd). In the former case, there can be multiple fixed points, i.e.,
this is a necessary condition for multistability.
On the other hand, when there are an odd number of repressors, then $F_i(x)$ is positive and monotonically
decreasing, meaning that there is one and only one solution to the
fixed point equation $x^*_i=F_i(x_i^*)$. 
The system of equations (1) has one unique fixed
point, which we denote ${\bf x}^*$. To perform the stability analysis, we write the characteristic
polynomial evaluated at this point:
\begin{equation}
\prod_i \left[\lambda-\partial_x g_i(x,y)|_{x=x^*}\right]=\prod_i \partial_y g_i(x,y)|_{x=x^*}.
\end{equation}

The above equation can be greatly simplified using the relation
$F'(x)=\prod_i \partial_y g_i(x,y)/\partial_x g_i(x,y)$, which is a
consequence of the implicit function theorem and the chain rule.  One
then obtains the following equation:
\begin{equation}\label{linstab_s}
\prod_{i=1}^N\left(\frac{\lambda}{h_i}+1\right)=F'(x^*)
\end{equation}
where the $h_i=- \partial_x g_i(x_i,x_{i=1})|_{x^*}$ are the
degradation rates at the fixed point. Notice that, because $F'(x)$ is always
negative in a negative feedback loop, all coefficients of the characteristic polynomial
are non-negative, hence it can not have real positive
roots. This means that the destabilization of the fixed point can only
occur via a Hopf bifurcation, i.e. with two complex conjugate
eigenvalues crossing into the positive real half-plane.

\begin{figure}[ht]
\begin{center}
\includegraphics[width=7cm]{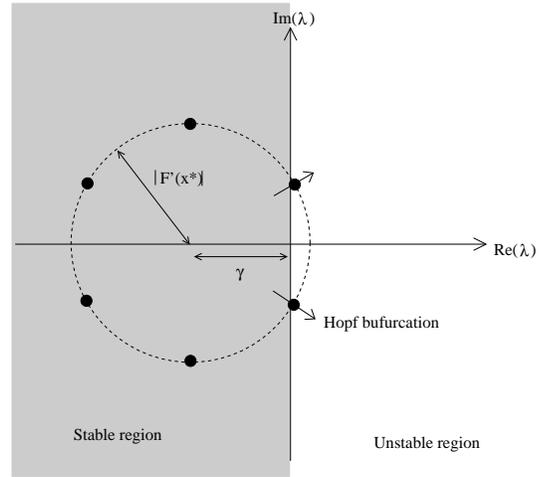}
\end{center}
\caption{Sketch of the Hopf bifurcation in the eigenvalue complex plane, 
in the case in which all the degradation rates are equal to a constant $\gamma$.\label{figure3_s}}
\end{figure}

In the simple case in which all the degradation
rates are equal and unchanging (i.e, $h_i=\gamma$, a constant) the roots of the polynomial (\ref{linstab_s}) in
the complex plane are the vertices of a polygon centered on $-\gamma$ with a
radius $|F'|$ as sketched in Fig. \ref{figure3_s}.
Therefore, the fixed point will remain stable as long as
\begin{equation}
|F'(x^*)|\cos(\pi/N)<\gamma.
\end{equation}
In this case, Hopf's theorem (see e.g. \cite{murray}) ensures the existence of a periodic orbit
close to the transition value, whose period is:
\begin{equation}
  \label{eq:hopfperiod_s}
  T=2\pi/Im(\lambda)
\end{equation}
which, in the simple case of equal degradation timescales, becomes
$T=2\pi/[|F'(x^*)|\cdot \sin(\pi/N)]$.
Notice that the Hopf theorem does not ensure that the orbit is stable;
however, since the system is bounded and there are no other fixed
points, we expect the orbit to be attracting, at least close to the
transition point. 

Now we apply these ideas to the 3-repressor example discussed in the main text:
\begin{equation}\label{simpleexample_s}
\frac{dx_i}{dt}=c-\gamma x_i+\alpha\frac{1}{1+(x_{i-1}/K_i)^h} \qquad i=1\ldots 3.
\label{3cycle_eq_s}
\end{equation}
The coordinates of the fixed point are all equal due to symmetry. We denote by
$x^*$ the solution to the equation $\gamma x=c+\alpha/(1+(x/K)^h)$. Then the
characteristic polynomial is simply:
\begin{equation}
(\lambda+\gamma)^3=-\left(\frac{\alpha}{1+(x^*/K_i)^h}\right)^3
\end{equation}
Notice, that this can be written as:
\begin{equation}
\left(\frac{\lambda}{\gamma}+1\right)^3=F'(x^*)
\end{equation}
The stability condition is then:
\begin{equation}
|F'(x^*)|\cos(\pi/3)<\gamma.
\end{equation}

\begin{figure}
(a)\hfill(b)\hfill ~\\
\begin{center}
\includegraphics[width=4.5cm]{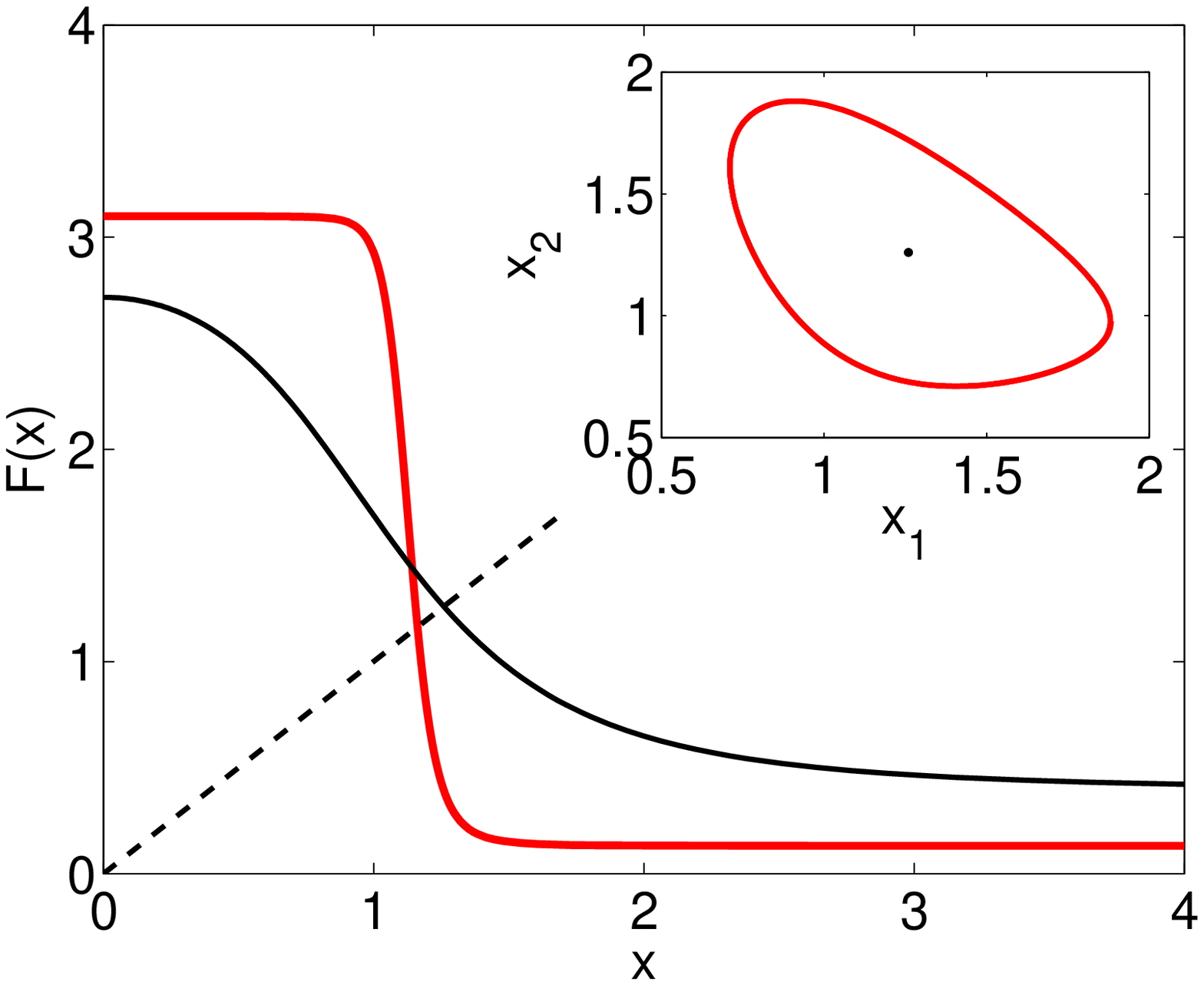}
\includegraphics[width=4.5cm]{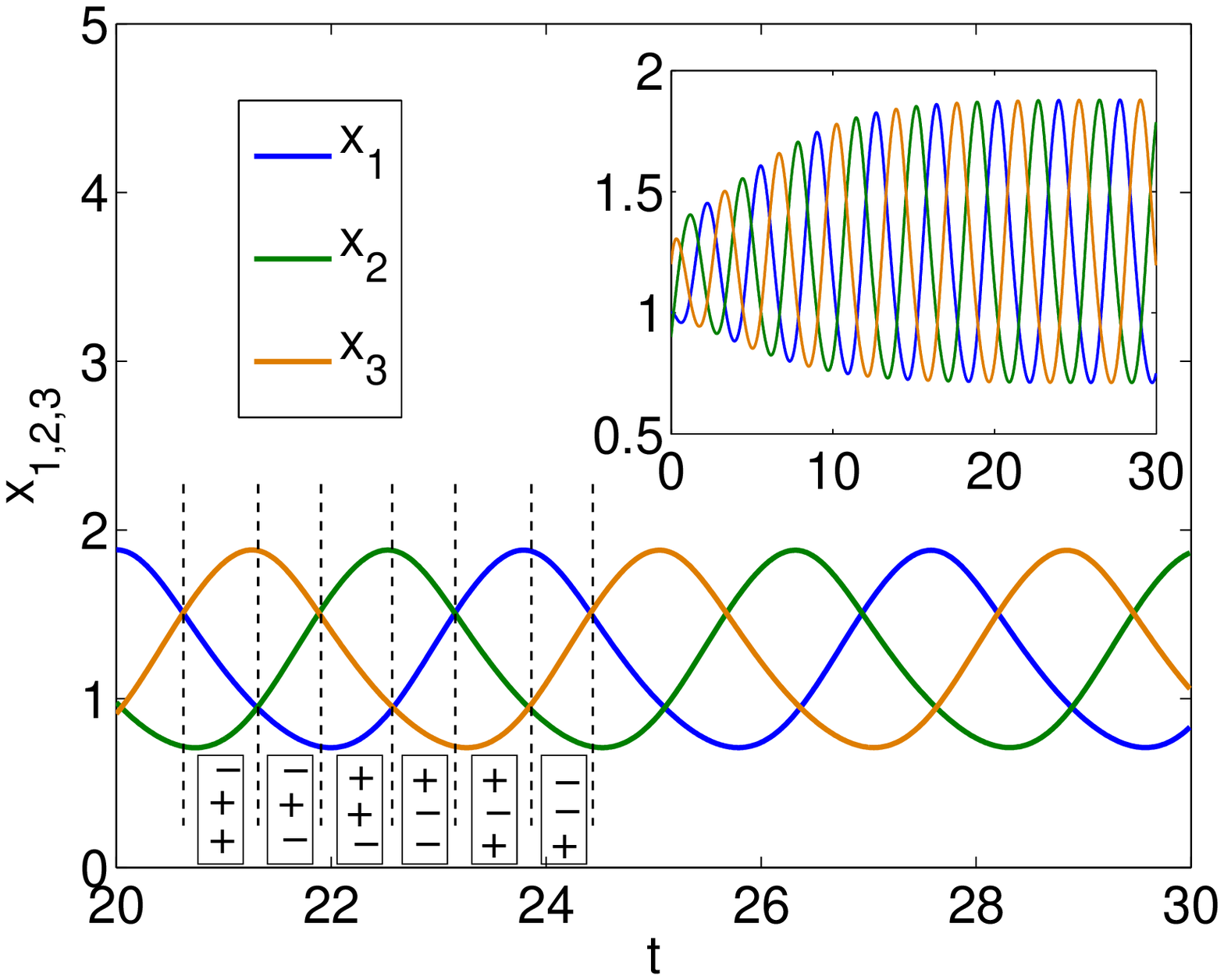}
\end{center}
\caption{A simple genetic oscillator described by eq. (\ref{simpleexample_s}).
(a) Plot of $F_1(x)$ for $h=2$ (black solid curve) and $h=4$ (red curve). Other parameters are kept fixed: $\alpha=3.0,~c=0.1,~K=1,~\gamma=1$.
The fixed point value of $x_1$ in each case lies at the intersection of the curves
with the dotted line.
Inset shows the corresponding trajectories
in the $x_1x_2$ plane.
(b) Time evolution of $x_{1,2,3}$ for the $h=4$ case after a stable limit cycle is
reached (inset shows a longer time plot including the transient). Also shown is
the symbolic dynamics for this time evolution.\label{examplefig_s}}
\end{figure}

Fig. \ref{examplefig_s}a shows that $|F'|$ satisfies this condition for
$h=2$ (black curve) but not for $h=4$ (red curve), when the other parameters
are kept fixed at the values $\alpha=3.0,~c=0.1,~K=1,~\gamma=1$.  Consequently, for
$h=2$ the trajectory converges to a stable fixed point, whereas it
converges to a stable limit cycle for $h=4$, as shown in the inset of
Fig. \ref{examplefig_s}a.
                                                                                                                                                                                                                                             

\section*{Symbolic dynamics}
In this section we give the mathematical details of the section
``Symbolic dynamics'' in the paper. This section is organized in the same
order as in the main text and we will repeat
the statements made there in a mathematically more rigorous way.

Let us use $\Gamma$ to denote the phase space, i.e. the positive orthant:
\begin{equation}
\Gamma= \{x_i>0\} \qquad \forall i=1\ldots N.
\end{equation}
 In the systems in which trajectories are bounded only when the
concentrations do not grow more than some maximum value (this is the case of
saturated degradation), all the following considerations still hold, with
the prescription of taking $\Gamma$ to be the subset of the positive orthant
in which the concentrations are bounded.  

Our goal is to describe how the
space $\Gamma$ is partitioned by the $N$ nullclines $\mu_i$ defined by
$g_i(x_i, x_{i-1})=0$. The properties we are about to state are all
consequences of the monotonicity of the functions $g_i(x_i, x_{i-1})=0$, the
constraint of having bounded and persistent orbits and the existence of a
unique fixed point $\mathbf{x}^*$.

It is worth remarking here
that the existence of a fixed point is more a consequence of the boundedness
condition, rather than the monotonicity condition. Indeed, if the
function $f_i(x_{i-1})$ and $f_{i-1}^{-1}(x_{i-1})$ (see Eq.(2))
have independent support, they will obviously have no
intersection. But in this case the system will not be persistent:
persistence requires that $\lim_{x\rightarrow0}g(x,y)>0\ \forall y$ and
boundedness requires that $\lim_{x\rightarrow\infty}g(x,y)<0\ \forall y$; in
this case, the nullclines have to cross.  This fact is crucial also for the
other considerations in this section on the phase space portrait.  Actually
the existence of a fixed point could have been demonstrated using only the
boundedness and persistence hypothesis by means of Brouwer's fixed point
theorem (see e.g. \cite{hofbauer}). 

\begin{figure}[hbt]
\begin{center}
\includegraphics[width=7cm]{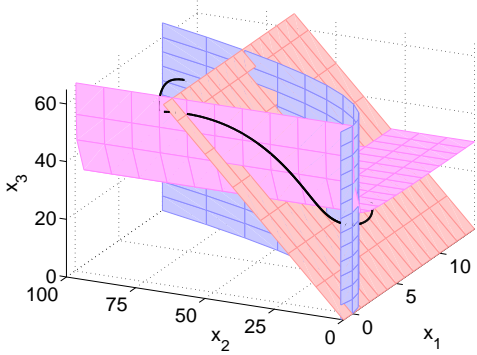}
\end{center}
\caption{Stable periodic orbit and plot of the nullclines for a model
of the p53 system, defined by the system of equations (\ref{modellop53_s}).\label{sectorsp53_s}}
\end{figure}

Returning to the partitioning of $\Gamma$, the first important property is
that any nullcline divides $\Gamma$ into two simply connected sets, one in
which $g_i(\mathbf{x})>0$ and one in which $g_i(\mathbf{x})<0$. Notice also
that these manifolds cannot be tangent at the fixed point because of 
monotonicity and since they can depend on at most one common variable.  
All these properties imply that the $\Gamma$ space is partitioned by the
nullclines into $2^N$ simply connected subsets, which we called ``sectors" in
the main text. In each of these sectors every component of the vector field
has a definite, unchanging, sign. We use here the same notation as the main text
and denote each of these sectors with a sign vector like $(+,-,-....+)$. As
stated in the paper, the uniqueness of the fixed point and the fact that the
field has a constant sign inside a sector allows one to exclude the
possibility of an attractor entirely contained within a sector.  In the rest
of this section, we discuss the case in which the fixed point is
unstable. Since we assume that trajectories are bounded, starting from a
sector the trajectory has to leave it by crossing one of the $N$
boundaries\footnote{Strictly speaking, one has to exclude the possibility
that the trajectory leaves the sector by crossing at the intersection between two
nullclines, i.e. one of the sets $\mu_i\cap\mu_j$. This would correspond to
two components of the field changing sign at the same time. We will not
consider this case here since it not robust, occurring only for a set
of parameter values that is of measure zero.}. 
We show in Fig. 3 in the main text that a given boundary $g_i=0$ can be
crossed in just one direction using a simple, two dimensional example. We show
here a three dimensional example of how a stable periodic orbit crosses the
nullclines. We consider the following model, consisting of
one repressor and two activators (this is similar to one of the
3-variable models of the p53 system discussed in \cite{alonp53s}):
\begin{eqnarray}\label{modellop53_s}
dx_1/dt&=&s-x_3x_1/(K+x_1)\nonumber\\
dx_2/dt&=&x_1^2-x_2\nonumber\\
dx_3/dt&=&x_2-x_3
\end{eqnarray}
This system of equations has a stable periodic orbit as an attractor for parameters values
$s=30$, $K=.1$. The phase space portrait, together with a plot of the
nullclines, is shown in Fig.(\ref{sectorsp53_s}).

Indeed, due to the fact that a given nullcline $g_i=0$ is ``flat'' in all directions
perpendicular to $x_i$ and $x_{i-1}$, no new topological features
appear in the higher dimensional cases; in particular the direction in
which this nullcline can be crossed depends only on the sign of $g_i$ and
$g_{i-1}$, and never on any other nullcline.  This directly follows from the fact that
all the manifolds $\mu_i\cap\mu_j$ (intersections of
a pair of nullclines) are simply connected; it is easy use the function
$f_i(x)$ defined in Eq.(2) to write an explicit and
continuous parameterization of these manifolds.
It should be clear at this point that the following fact is true in any dimension $N$:
{\em the portion of
the nullcline which forms the boundary between two adjacent sectors can only be
crossed in one direction}. 

The above italicised statement directly implies the two transition rules stated in the main text,
which we repeat here:
\begin{itemize}
\item If the variable $(i-1)$ represses $i$, the nullcline $i$
  can be crossed if $g_i$ and $g_{i-1}$ have the same sign.
\item If the variable $(i-1)$ activates $i$, the nullcline $i$
  can be crossed if $g_i$ and $g_{i-1}$ have opposite signs.
\end{itemize}

\begin{figure}[ht]
\begin{center}
\includegraphics[width=7cm]{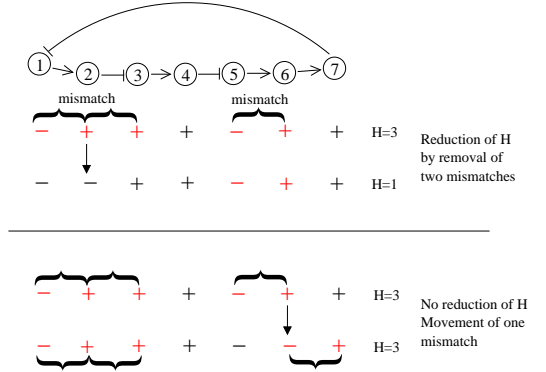}
\end{center}
\caption{Consequences of crossing a nullcline for the sign vector and the
  quantity $H$. Upper panel: a sign is changed between two mismatches and,
  consequently, $H$ decreases by 2. Lower panel: a sign is changed next to just
  one mismatch. In this case, the mismatch simply moves one step ahead in
  the loop; the value of $H$ does not change.\label{figureH_s}}
\end{figure}

We associate to a given symbol, or sector, the quantity $H$ defined as the number
of boundaries that can be crossed from that sector. Notice that $H=0$ is
impossible. For this to happen, the above rules must be violated by every adjacent pair
of signs, i.e., the signs on both ends of an activation arrow must be the same, while the
two signs on both ends of a repression arrow must be different. As there are an odd number of
repressors, this is impossible. Therefore, $1\le H\le N$.
When the trajectory crosses a nullcline $g_i=0$, 
as a simple consequence of the transition rules, $H$ either stays
constant if the nullcline $g_{i+1}$ can be crossed
from the new sector, or decreases by two if $g_{i+1}$ cannot be
crossed (see Fig. \ref{figureH_s}). 
Physically, if we think of a crossable boundary as an
unsatisfied bond between sign $i$ and $i-1$ (termed ``mismatches" in the main text), 
$H$ is the number of such mismatches, hence it quantifies the level of ``frustration'' in the system. The time
evolution can then (i) solve two neighboring unsatisfied bonds, or (ii)
shift an unsatisfied bond one place to the right, i.e. from $i$ to $i+1$.
As a consequence: $H$ can never increase, and it must always be an odd number.

$i-1$ (see Fig \ref{figureH_s}).
We expect the system to end up in a state in which $H=1$ and the
unsatisfied bond keeps on moving around the loop in the direction of the arrows. This represents, at the
level of symbolic dynamics, a single ``signal" traveling around the loop. A
direct implication is that the extremal points (maxima or minima) of
all variables should appear in the time series in the order in which
the species are arranged in the cycle.

This is the simplest scenario and is the only one for $N<4$. What can
actually happen for $N\ge 4$ is that $H=3$ can become a stable state,
with $3$ different signals traveling along the feedback loop. 
However, even if there is an attractor with $H=3$, it must anyway
coexist with an $H=1$ attractor, since if a trajectory ever starts from, or enters, a
sector with $H=1$, it can never return to an $H=3$ sector.
Further, the $H=3$ attractor is likely to require some fine tuning of parameters to avoid
one mismatch travelling ``faster", catching up and annihilating another one.
It is thus likely that any
perturbation or noise would bring the system to the ``ground state''
attractor characterized by $H=1$, and that is the one we expect to
observe in biological systems.

\section*{Unobserved variables}
We conclude by arguing that the presence of unobserved variables
 does not alter our conclusions. Essentially the rules we stated
 depend only on the overall sign between two variables and it is
 irrelevant if there are species in between mediating their
 interaction. This can be easily shown by considering all the possible
 cases of an intermediate species being an
 activator/repressor and being activated/repressed. Then it can be
 generalized by induction to an arbitrary number of mediators. Since
 the general demonstration is straightforward, consisting essentially
 of considering all the possible cases, we show here just one example:
 the three-repressor loop of the previous section, in which one of
 the species is unobserved. We write its symbolic dynamics (as in Fig. \ref{examplefig_s}b) by
 means of the rules we shown and simply cancel the second row:
\begin{eqnarray}
&\begin{array}{c}x_1\\x_2\\x_3\end{array}
\left(\begin{array}{c}-\\+\\-\end{array}\right)
\left(\begin{array}{c}+\\+\\-\end{array}\right)
\left(\begin{array}{c}+\\-\\-\end{array}\right)
\left(\begin{array}{c}+\\-\\+\end{array}\right)
\left(\begin{array}{c}-\\-\\+\end{array}\right)
\left(\begin{array}{c}-\\+\\+\end{array}\right)\nonumber\\
&\longrightarrow
\left(\begin{array}{c}-\\-\end{array}\right)
\left(\begin{array}{c}+\\-\end{array}\right)
\left(\begin{array}{c}+\\+\end{array}\right)
\left(\begin{array}{c}-\\+\end{array}\right)
\begin{array}{c}x_1\\x_3\end{array}
\end{eqnarray}
The resulting symbolic dynamics is that 
 of a two species loop with one activator and one repressor (like the p53-Mdm2 oscillations
in Fig. 1 in the main text). Here, as expected, two
 repressing links become one activating link if the intermediate species, $x_2$, is
 unobserved.

\end{document}